# Spin Transfer Torque Generated by the Topological Insulator Bi$_2$Se$_3$

A. R. Mellnik,[1] J. S. Lee,[2] A. Richardella,[2] J. L. Grab,[1] P. J. Mintun,[1] M. H. Fischer,[1,3] A. Vaezi,[1] A. Manchon,[4] E.-A. Kim,[1] N. Samarth,[2] and D. C. Ralph[1,5]

[1] Cornell University, Ithaca, NY 14853

[2] Department of Physics, The Pennsylvania State University, University Park, PA 16802

[3] Weizmann Institute of Science, Rehovot 76100, Israel

[4] King Abdullah University of Science and Technology (KAUST), Physical Sciences and Engineering Division, Thuwal 23955-6900, Saudi Arabia

[5] Kavli Institute at Cornell, Ithaca, NY 14853

**Magnetic devices are a leading contender for implementing memory and logic technologies that are nonvolatile, that can scale to high density and high speed, and that do not suffer wear-out. However, widespread applications of magnetic memory and logic devices will require the development of efficient mechanisms for reorienting their magnetization using the least possible current and power[1]. There has been considerable recent progress in this effort, in particular discoveries that spin-orbit interactions in heavy metal/ferromagnet bilayers can yield strong current-driven torques on the magnetic layer[2-11], via the spin Hall effect[12,13] in the heavy metal or the Rashba-Edelstein effect[14,15] in the ferromagnet. As part of the search for materials to provide even more efficient spin-orbit-induced torques, some proposals[16-19] have suggested topological insulators (TIs)[20,21], which possess a surface state in which the effects of spin-orbit coupling are maximal in the sense that an electron's spin orientation is locked relative to its propagation direction. Here we report experiments showing that charge current flowing in-plane in a thin film of the TI $Bi_2Se_3$ at room temperature can indeed apply a strong spin-transfer torque to an adjacent ferromagnetic permalloy (Py = $Ni_{81}Fe_{19}$) thin film, with a direction consistent with that expected from the topological surface state. We find that the strength of the torque per unit charge current density in the $Bi_2Se_3$ is greater than for any other spin-torque source material measured to date, even for non-ideal TI films wherein the surface states coexist with bulk conduction. Our data suggest that TIs have potential to enable very efficient electrical manipulation of magnetic materials at room temperature for memory and logic applications.**

The proposed mechanism[16-19] that motivates our study of TIs as sources of current-induced spin torque is illustrated in Fig. 1. When an in-plane current flows in the surface state of a TI, more forward-going electron states are occupied than backward-going states and, because of the helical spin-momentum locking of the surface state, this necessarily means that the flow of charge is accompanied by a non-equilibrium surface spin accumulation with the spin moment in the $-\hat{x}$ direction as depicted in Fig. 1. If this spin accumulation couples to an adjacent magnetic film, the resulting flow of spin angular momentum will exert a spin-transfer torque on the magnet. It is important to note that this mechanism is not related to any physics near the Dirac point of the TI, which might be disrupted by coupling to a ferromagnet, and does not depend on having zero bulk conductivity within the TI. This torque is related to the Rashba-Edelstein effect in non-topological materials[14]. However, the helical spin-momentum locking of the TI surface state produces a different sign and a much larger magnitude of spin accumulation compared to non-topological materials (see Supplementary Information).

Our samples are bilayers consisting of 8 nm of $Bi_2Se_3$ and 8 or 16 nm of Py, with an oxidized aluminum cap to prevent oxidation of the Py surface, patterned into strips 10-80 μm long and 2.5-24 μm wide. The $Bi_2Se_3$ is grown by molecular beam epitaxy (MBE) and the Py by sputtering (see Methods). This sample geometry is actually not ideal for measuring torque due to current flow within the $Bi_2Se_3$, because the average resistivity of the $Bi_2Se_3$ is 25 times or more that of the metallic Py so that the great majority of the current is shunted through the Py and does not contribute to the torque. Nevertheless, the torque from the $Bi_2Se_3$ is still strong enough to be measured accurately.

We determine the strength of current-induced torque with a spin-torque ferromagnetic resonance (ST-FMR) technique developed previously to measure the spin Hall torque from heavy metals[4,5]. Using the circuit shown in Fig. 2, we apply a microwave current with fixed frequency and sweep an in-plane magnetic field through the ferromagnetic resonance condition.

The oscillating current-induced torque causes the Py magnetization to precess, yielding resistance oscillations on account of the anisotropic magnetoresistance (AMR) of Py. We measure the resonance lineshape through a direct voltage $V_{mix}$ that results from mixing between the applied alternating current and the oscillating resistance. The two vector components of the current-induced torque, in the $\hat{m} \times (\hat{x} \times \hat{m})$ ($\parallel$, "in-plane") and $\hat{x} \times \hat{m}$ ($\perp$, "perpendicular") directions (see Fig. 2a) are obtained from the amplitudes of the symmetric and antisymmetric components of the resonance lineshape (see Methods).

Fig. 3a shows the results of the ST-FMR measurement for a 50 μm × 15 μm device with 8 nm Bi$_2$Se$_3$ / 16 nm Py and with an in-plane magnetic field oriented φ = 45° relative to the current. The quality of the theoretical fit (to Eq. (2) in Methods) is excellent, from which we can determine that the oscillating torque per unit moment on the Py induced by the in-plane current has the components $\tau_\parallel = (2.7 \pm 0.3) \times 10^{-5}$ T and $\tau_\perp = (3.7 \pm 0.4) \times 10^{-5}$ T. The dependence of both the symmetric and antisymmetric components of $V_{mix}$ on φ is to good accuracy $V_{mix} \propto \cos(\varphi)^2 \sin(\varphi)$ (Fig. 3b). This is as expected for the ST-FMR signal [Eq. (2) in Methods], since for the AMR resistance (R) one has $dR/d\varphi \propto \cos(\varphi)\sin(\varphi)$ and the spin torques arising from spin accumulation should give $\tau_\parallel, \tau_\perp \propto \cos(\varphi)$.

For comparisons with first-principles calculations, the quantities of primary interest are the effective spin current conductivities,[22] $\sigma_{S,i} \equiv J_{S,i} / E = \tau_i M_s t_{mag} / [E \cos(\varphi)]$ where $J_{S,i}$ is the $i$-component ($\parallel$ or $\perp$) of the spin current density absorbed by the ferromagnet for φ = 0°, E is the amplitude of the electric field, and $M_s t_{mag}$ is the saturation magnetization times the thickness of the magnetic layer. For the measurement in Fig. 3a, the microwave field is $E = (0.8 \pm 0.1$ V$)/(50$ μm$)$ and $M_s t_{mag} = 14.2$ mA from vibrating sample magnetometry on test films. Averaging over multiple devices with 8 nm Bi$_2$Se$_3$ / 16 nm Py and frequencies 6-10 GHz, we find that

$\sigma_{S,\parallel} = (1.1 \pm 0.2) \times 10^5 \, \hbar/(2e) \, (\Omega m)^{-1}$ and $\sigma_{S,\perp} = (1.4 \pm 0.2) \times 10^5 \, \hbar/(2e) \, (\Omega m)^{-1}$.

Performing the same measurement on 8 nm Bi$_2$Se$_3$ / 8 nm Py devices, we find $\sigma_{S,\parallel} = (2.0 \pm 0.4) \times 10^5 \, \hbar/(2e) \, (\Omega m)^{-1}$ and $\sigma_{S,\perp} = (1.6 \pm 0.2) \times 10^5 \, \hbar/(2e) \, (\Omega m)^{-1}$. These values are comparable to the spin-current conductivities for the most efficient spin current sources known previously, heavy metals which generate current-induced spin currents by the spin Hall effect[4,5,23,24] (see Table 1). Since the electrical conductivities of the heavy metals are much greater than Bi$_2$Se$_3$, this is a first indication that the strength of spin torque per unit current is greater in Bi$_2$Se$_3$ (see below). There is some potential uncertainty in our determination of $\sigma_{S,\parallel}$ in that pumping of spins from the precessing ferromagnet into the Bi$_2$Se$_3$ together with the inverse Edelstein effect could produce an additional contribution to the symmetric ST-FMR signal that is not accounted for in our analysis. However, for this effect to be significant to our measurements would require a value of $|\sigma_{S,\parallel}|$ as large as or larger than we determine from the ST-FMR analysis (see Supplementary Information). Therefore, the presence of a spin-pumping contribution would not change our central conclusion that Bi$_2$Se$_3$ provides very large values of $\sigma_{S,\parallel}$ and $\sigma_{S,\perp}$.

The sign we measure for $\sigma_{S,\parallel}$ is consistent with expectations for spin transfer from the current-induced spin accumulation in the Bi$_2$Se$_3$ TI surface state with a chemical potential above the Dirac point, i.e. spin angular momentum in the direction $-\hat{z} \times \hat{k}$ where $\hat{z}$ is the surface normal and $\hat{k}$ the electron wavevector direction[25,26], or spin moment in the direction $\hat{z} \times \hat{k}$ (Fig. 1). The sign of $\sigma_{S,\perp}$ is the same as the torque due to an Oersted field, but the magnitude is much larger (see below). Control experiments on Py layers without Bi$_2$Se$_3$ and Py/Pt bilayers yield much smaller values of $\sigma_{S,\parallel}$ and $\sigma_{S,\perp}$ (see Supplementary Information).

An independent measurement confirming the value of $\sigma_{S,\perp}$ can be obtained by measuring the current-induced shift of the ST-FMR resonant field (see Supplementary Information). Current-induced changes in the ST-FMR resonance width have been used previously as an alternative method to measure $\sigma_{S,\parallel}$[2,4], but we find that this is not possible in our samples because the resonance width is not a linear function of frequency (see Supplementary Information).

For applications, the figure of merit of primary interest is generally the "spin torque angle" $\theta_\parallel$, the strength of the in-plane component of torque per unit applied charge current density $J$ in the spin-current source material [$\theta_\parallel \equiv (2e/\hbar)J_{S,\parallel}/J = (2e/\hbar)\sigma_{S,\parallel}/\sigma$; here $\sigma$ is the charge conductivity in the spin-current source material], because this quantity fundamentally determines the current needed for efficient magnetic manipulation[5,6,10,11]. Unlike $\sigma_{S,\parallel}$ and $\sigma_{S,\perp}$ which depend directly on the easily-measured electric field, determining $\theta_\parallel$ requires knowing the average value of $\sigma$ in $Bi_2Se_3$ when it is in contact with the Py. This is tricky because the addition of Py onto $Bi_2Se_3$ causes band bending in the $Bi_2Se_3$ that increases $\sigma$ relative to $Bi_2Se_3$ in isolation[26-28]. To mimic the band-bending effects of Py, we have measured test samples of $Bi_2Se_3$ with an insulating $Al_2O_x$ cap, made by depositing 2 nm of Al and oxidizing in a dry $O_2/N_2$ mixture, that also shifts the chemical potential up. For 8 nm of $Bi_2Se_3$ capped with $Al_2O_x$ we find at room temperature an average 2D charge density of $9.4 \times 10^{13}$ cm$^{-2}$ by Hall measurements and an average conductivity $5.7 \times 10^4$ $(\Omega m)^{-1}$. Valla *et al.* used ARPES to measure the band bending of $Bi_2Se_3$ when coupled to various metals including Ni and Rb[28]. The band bending due to Rb was stronger than for Ni and corresponded to a maximal doping of $5 \times 10^{13}$ cm$^{-2}$. We therefore make the rough estimate that our Py = $Ni_{81}Fe_{19}$ should increase the average conductivity of the Bi$_2$Se$_3$ by no more than the value we measure for the Al$_2$O$_x$ cap. This yields a lower bound $\theta_\parallel \geq 3.5$ based on 8 nm Py devices (16 nm Py devices give a lower bound of $\theta_\parallel \geq 2.0$). This is the largest spin torque ratio measured for any spin-source material (Table 1).

The value that we determine for the out-of-plane spin current conductivity, $\sigma_{S,\perp} \approx 1.5 \times 10^5 \, \hbar/(2e) \, (\Omega m)^{-1}$, is much larger than can be explained by the Oersted field. Given our estimates of $\sigma$ for Bi$_2$Se$_3$ after band bending by the Py, the out-of-plane spin-current conductivity that would be generated by the Oersted field alone is just $\sigma_{\perp,Oe} \approx 6.1 \times 10^3 \, \hbar/(2e) \, (\Omega m)^{-1}$ for the 8 nm Py samples and $\sigma_{\perp,Oe} \approx 1.2 \times 10^4 \, \hbar/(2e) \, (\Omega m)^{-1}$ for the 16 nm Py samples, smaller by factors of 23 and 13, respectively, than our measurements.

Our findings are in excellent agreement with a model that considers non-equilibrium spin accumulation near the Bi$_2$Se$_3$ surface and its diffusion into the Py layer (see Supplementary Information). This model predicts correctly the signs of both $\sigma_{S,\parallel}$ and $\sigma_{S,\perp}$, while torque due to the Rashba-Edelstein effect from non-topological interface states would give opposite signs. Furthermore our model predicts in agreement with the experiment that the magnitude of $\sigma_{S,\perp}$ driven by the TI surface state should have a magnitude comparable to $\sigma_{S,\parallel}$, and much larger than an Oersted field effect.

Our findings have potential importance for technology in that the spin torque angle for Bi$_2$Se$_3$ at room temperature is larger than for any previously-measured spin-current source material. However, as noted above, for practical applications the specific layer structure of our devices -- TI/metallic magnet -- does not make good use of this high intrinsic efficiency because most of the applied current is shunted through the metallic magnet and does not contribute to spin-current generation within the TI. Applications will likely require coupling TIs to insulating (or high resistivity) magnets so that the majority of current will flow in the TI. Using insulating magnets may also have an advantage of providing much lower values of Gilbert damping than metallic magnets[29], which can yield an additional reduction in the current levels needed for manipulation via the spin-torque anti-damping mechanism. Additionally, using an insulating magnet would allow electrostatic gating of bilayer devices, thus allowing the chemical potential to be tuned to eliminate any deleterious effects of bulk conduction. Our results therefore point toward a new strategy for implementing low-power nonvolatile magnetic memory and logic structures, using TIs as room-temperature sources of spin torque coupled to insulating magnetic layers, to achieve potential switching efficiencies better than any other known mechanism.

**Methods**

*Bilayer deposition and electrical properties.* The $Bi_2Se_3$ thin films were grown by MBE on 2-inch C-axis oriented epi-ready sapphire wafers at a substrate temperature of approximately 300 C using thermal evaporation of high purity (5N) elemental Bi and Se from Knudsen cells with a Se to Bi beam equivalent pressure ratio of approximately 14:1. The growth rate was 0.8 quintuple-layers per minute. After growth the films were capped with ~ 4 nm of Se to protect the surface. The $Bi_2Se_3$ film thickness was verified by x-ray reflection and the films were characterized by atomic force microscopy and high resolution x-ray diffraction, yielding rocking curve widths less than 0.15°. The samples were then transferred via air to a magnetron sputtering chamber with a base pressure of $2 \times 10^{-9}$ Torr, where the Se capping layer was removed by heating the sample to 240 C for one hour. After cooling to room temperature, Py was sputtered on the bare $Bi_2Se_3$ followed by a 2 nm Al capping layer, which was subsequently oxidized in air. The average resistivity of the Py layer changes slightly with thickness due to the roughness of the $Bi_2Se_3$, and is 71.7 μΩcm for a nominally 8 nm layer and 45.9 μΩcm for a nominally 16 nm layer. The average carrier density in the Se-capped $Bi_2Se_3$ was $2.64 \times 10^{19}$ $cm^{-3}$ and n-type as determined by Hall-effect measurements (average areal density of $2.1 \times 10^{13}$ $cm^{-2}$), so that the chemical potential lies above the Dirac point, at the edge of the conduction band. With the removal of the Se and the deposition of Py we anticipate a shift of the chemical potential even higher into the conduction band[26-28].

To make devices, we pattern the $Bi_2Se_3$/Py bilayers using optical lithography and ion milling, with electrical contacts made from 3 nm Ti / 150 nm Pt in a symmetric geometry (Fig. 2b) so that when the samples are contacted using a ground-signal-ground high-frequency probe the currents traveling in the contacts do not produce a net Oersted field acting on the sample.

*Analysis of ST-FMR measurements.* We interpret the ST-FMR signals within a macrospin approximation for the magnetization direction $\hat{m}$ using the Landau-Lifschitz-Gilbert-Slonczewski equation of motion[30]

$$\frac{d\hat{m}}{dt} = -\gamma\hat{m}\times\left(\vec{B}_{ext} - \mu_0 M_{eff} m_z \hat{z}\right) + \alpha\hat{m}\times\frac{d\hat{m}}{dt} + \gamma\tau_\parallel \frac{\hat{m}\times(\hat{x}\times\hat{m})}{|\hat{x}\times\hat{m}|} + \gamma\tau_\perp \frac{\hat{x}\times\hat{m}}{|\hat{x}\times\hat{m}|}. \quad (1)$$

Here $\gamma$ is the absolute value of the gyromagnetic ratio, $\vec{B}_{ext}$ is the applied magnetic field, $\mu_0 M_{eff}$ is the out-of-plane demagnetization field, $\alpha$ is the Gilbert damping constant, $\tau_\parallel$ is the "in-plane" component of the current-induced torque per unit moment (the symmetry analogous to a spin Hall torque), $\tau_\perp$ is the perpendicular component (analogous to the torque due to an Oersted or Rashba field), and $\hat{x}$ and $\hat{z}$ are defined as shown in Fig. 2a. We calculate that near zero bias current and for small-angle precession the ST-FMR mixing voltage has the form

$$V_{mix} = -\frac{I_{RF}\gamma}{4}\left(\frac{dR}{d\varphi}\right)\left[\tau_\parallel \frac{1}{\Delta} F_S(B_{ext},\omega) + \tau_\perp \frac{[1+(\mu_0 M_{eff}/B_{ext})]^{1/2}}{\Delta} F_A(B_{ext},\omega)\right], \quad (2)$$

where $I_{RF}$ is the total microwave current flowing through the device, $R(\varphi)$ is the anisotropic magnetoresistance as a function of the in-plane magnetization angle $\varphi$, $\Delta = \alpha\gamma(2B_{ext} + \mu_0 M_{eff})/2$ is the zero-current linewidth, $F_S(B_{ext},\omega) = (2\Delta)^2\omega^2/[(\omega^2 - \omega_0^2)^2 + (2\Delta)^2\omega^2]$ and $F_A(B_{ext},\omega) = [(\omega_0^2 - \omega^2)/(2\omega\Delta)]F_S(B_{ext},\omega)$ are approximately symmetric and antisymmetric resonance lineshapes as a function of $B_{ext}$, $\omega$ is the microwave frequency, and $\omega_0 \equiv \gamma\sqrt{B_{ext}(\mu_0 M_{eff} + B_{ext})}$ is the resonance frequency. These expressions are equivalent to those in ref. 4 except that we no longer make the approximations that $\Delta \ll B_{ext}$ and $B_{ext} \ll \mu_0 M_{eff}$. The two components of the current-induced torque can therefore be determined from the amplitudes of the symmetric and antisymmetric components of the resonance. Spin pumping together with the inverse Edelstein effect can also provide an additional symmetric component to the resonance signal not accounted for here -- see the discussion in the Supplementary Information.

For the ST-FMR analysis, we calibrate the anisotropic magnetoresistance $R(\varphi)$ for each device by rotating a 0.07 T magnetic field within the sample plane using a projective-field magnet (Supplementary Information). We use a network analyzer to calibrate the transmission coefficients of our microwave circuit and the reflection coefficient of each sample. Based on this calibration, we determine $I_{RF}$ and the strength of the electric field in the device for a given microwave power. The values we find from the ST-FMR measurements for $\sigma_{S,\parallel}$ and $\sigma_{S,\perp}$ are consistent for different applied microwave powers, demonstrating that the data correspond to the linear regime of small-angle precession, and they have no systematic variation with the length or width of the sample.

**Acknowledgements** We thank Bob Buhrman, Chi-Feng Pai, Neal Reynolds, and Jonathan Gibbons for discussions. Work at Cornell and Penn State was supported by DARPA (N66001-11-1-4110). We acknowledge additional funding for work at Cornell from the Army Research Office (W911NF-08-2-0032) and the NSF (DMR-1010768) and for work at Penn State from the Office of Naval Research (N00014-12-1-0117). ARM acknowledges a DOE Office of Science graduate fellowship and JLG acknowledges an NSF graduate fellowship. JSL acknowledges support from DARPA C-SPIN. This work was performed in part at the Cornell NanoScale Facility and the Penn State Nanofabrication Facility, both nodes of the National Nanotechnology Infrastructure Network (NNIN), which is supported by the NSF (ECS-0335765), and in the facilities of the Cornell Center for Materials Research, which is supported by the NSF/MRSEC program (DMR-1120296).

**Author Contributions** A.R.M., J.S.L., A.R., N.S., and D.C.R. conceived and designed the experiments. A.R.M., J.L.G., and P.J.M. performed the sample fabrication, measurements, and analysis. J.S.L, A.R., and N.S. developed the growth process for the $Bi_2Se_3$ layers. M.H.F, A.V., A.M., and E.A.K. performed theoretical modeling. N.S. and D.C.R. provided oversight and advice. A.R.M. and D.C.R. wrote the manuscript and all authors contributed to its final version.

Reprints and permissions information is available at www.nature.com/reprints

**Competing financial interests statement** The authors declare no competing financial interests.

Correspondence and requests for materials should be addressed to dcr14@cornell.edu.

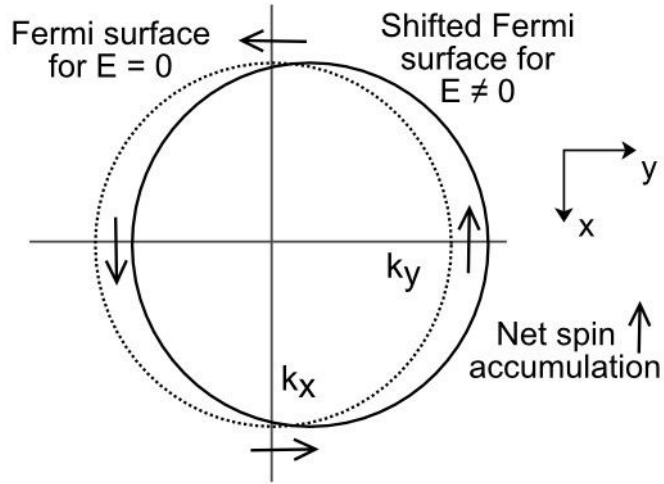

**Figure 1. Illustration of the mechanism by which an in-plane current in a topological insulator surface state generates a non-equilibrium surface spin accumulation, on account of locking between the spin direction and wavevector for electrons in the surface state.** The arrows denote the directions of spin magnetic moments, which are opposite to the corresponding spin angular momenta since the *g* factor of the electron is negative. For simplicity, the spins in this cartoon are depicted in the sample plane, although some canting out of plane is expected.

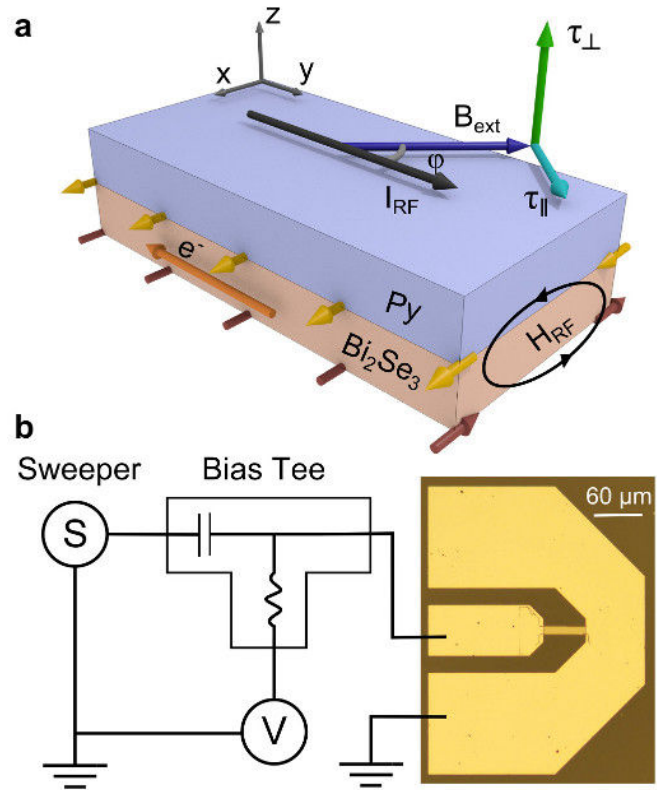

**Figure 2. Sample geometry used in the measurement**. **a,** Schematic diagram of the layer structure and coordinate system. The yellow and brown arrows denote spin moment directions. **b,** Depiction of the circuit used for the ST-FMR measurement and the sample contact geometry.

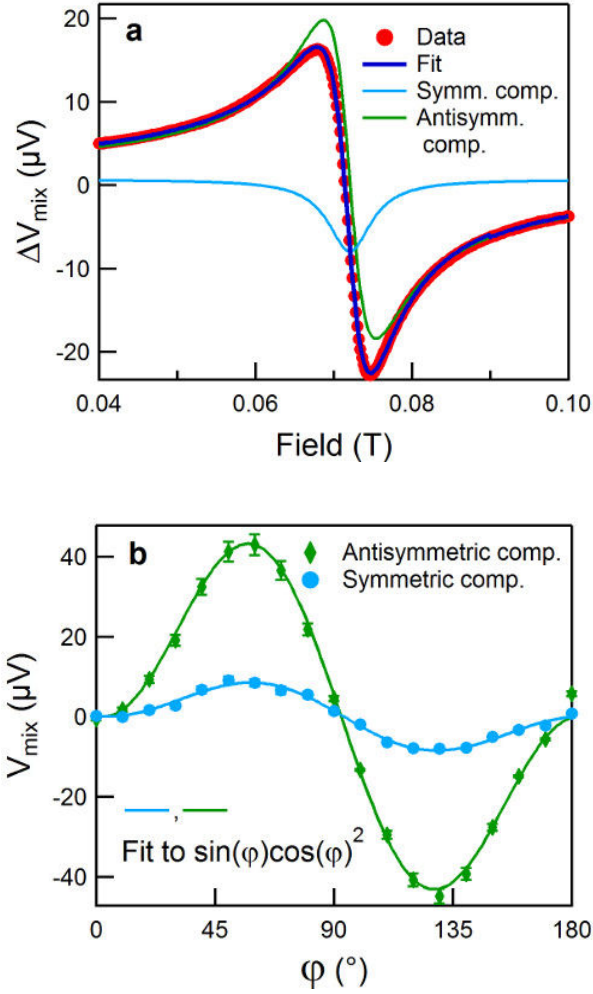

**Figure 3. ST-FMR measurements of the current-induced torque, with fits**. **a,** Measured ST-FMR resonance at room temperature with $\omega/2\pi = 8$ GHz for an 8 nm $Bi_2Se_3$/16 nm Py sample with dimensions 50 μm × 15 μm. A fixed microwave power of 5 dBm is absorbed by the device ($I_{RF} = 7.7 \pm 1.1$ mA) and $B_{ext}$ is oriented at an angle φ = 45° from the current direction. The lines are fits to Eq. (2) showing the symmetric and antisymmetric resonance components. **b,** Measured dependence on the magnetic field angle $\varphi$ for the symmetric and antisymmetric resonance components for a different sample (8 nm $Bi_2Se_3$/16 nm Py device, 80 μm × 24 μm). Experimental conditions in **b** are $\omega/2\pi = 9$ GHz, power absorbed = 6.9 dBm.

**Table 1. Comparison of the in-plane components of the spin current conductivity $\sigma_{S,\parallel}$ and the spin torque ratio $\theta_\parallel$ for Bi$_2$Se$_3$ at room temperature to other materials**. $\theta_\parallel$ is dimensionless and the units for $\sigma_{S,\parallel}$ are $10^5\ \hbar/(2e)\ (\Omega\mathrm{m})^{-1}$.

|  | Bi$_2$Se$_3$ (this work) | Pt (ref. 4) | β-Ta (ref. 5) | Cu(Bi) (ref. 23) | β-W (ref. 24) |
| --- | --- | --- | --- | --- | --- |
| $\theta_\parallel$ | 2.0-3.5 | 0.08 | 0.15 | 0.24 | 0.3 |
| $\sigma_{S,\parallel}$ | 1.1-2.0 | 3.4 | 0.8 | -- | 1.8 |

# Spin Transfer Torque Generated by the Topological Insulator $Bi_2Se_3$

## Supplementary Information

A. R. Mellnik,[1] J. S. Lee,[2] A. Richardella,[2] J. L. Grab,[1] P. J. Mintun,[1] M. H. Fischer,[1,3] A. Vaezi,[1] A. Manchon,[4] E.-A. Kim,[1] N. Samarth,[2] and D. C. Ralph[1,5]

[1] Cornell University, Ithaca, NY 14853

[2] Department of Physics, The Pennsylvania State University, University Park, PA 16802

[3] Weizmann Institute of Science, Rehovot 76100, Israel

[4] King Abdullah University of Science and Technology (KAUST), Physical Sciences and Engineering Division, Thuwal 23955-6900, Saudi Arabia

[5] Kavli Institute at Cornell, Ithaca, NY 14853

**I. Control experiments on single-layer Py samples and Pt/Py bilayers.**

As a first control experiment, we compared the ST-FMR signals measured for $Bi_2Se_3$/Py bilayers to single-layer Py films deposited on sapphire (Fig. S1a). As has already been reported previously[4], the ST-FMR signal for single-layer Py samples is not zero. For the single-layer Py we observe a small antisymmetric resonance that may be the result of nonuniform current density within the Py or the excitation of non-uniform magnetic modes. However, the signal from the single-layer Py is small enough to represent only a small correction to the $Bi_2Se_3$/Py results.

We also performed ST-FMR measurements on Pt/Py bilayers consisting of 6 nm sputtered Pt and 12 nm of evaporated Py, and the results for one such device are shown in Fig.

S1b. The resistivity of the Pt was 35 µΩ cm and $M_s t_{mag}$ = 5.0 mA for the evaporated Py films. Averaging over multiple devices and frequencies 6-10 GHz, we find $\sigma_{S,\parallel}^{Pt} = (1.9 \pm 0.4) \times 10^5$ $\hbar/(2e) (\Omega m)^{-1}$ and the spin-torque ratio $\theta_{\parallel}^{Pt} \approx 0.065$ which agrees with our group's previously reported values[4]. We also find that $\sigma_{S,\perp}^{Pt} = (2.4 \pm 0.5) \times 10^5$ $\hbar/(2e) (\Omega m)^{-1}$, which is comparable to the expected contribution from the Oersted field, $\sigma_{\perp,Oe}^{Pt} \approx 1.6 \times 10^5$ $\hbar/(2e) (\Omega m)^{-1}$.

## II. Measurements of $\sigma_{S,\parallel}$ based on the current dependence of the ferromagnetic resonance field

A direct current produces a small shift in the center field of the ST-FMR signal at fixed frequency (Fig. S2). This can be explained as due to a current-induced out-of-plane DC spin torque $\tau_{\perp,DC}$ which to a good approximation alters the ST-FMR signal equivalent to a shift in the applied magnetic field, $\vec{B}_{ext} \to \vec{B}_{ext} - \tau_{\perp,DC} \tan(\varphi)$. By this method we estimate $\sigma_{S,\perp} = (2.2 \pm 0.6) \times 10^5$ $\hbar/(2e) (\Omega m)^{-1}$ in 8 nm Bi$_2$Se$_3$/16 nm Py devices, which is consistent within the experimental uncertainty with our ST-FMR measurement in these devices, $\sigma_{S,\perp} = (1.4 \pm 0.2) \times 10^5$ $\hbar/(2e) (\Omega m)^{-1}$.

## III. Anomalous behavior of the ferromagnetic resonance linewidths

In heavy-metal/ferromagnet samples, the in-plane current-induced torque due to the spin Hall effect can be determined accurately using measurements of the current dependence of the FMR linewidth, since the in-plane torque can alter the effective Gilbert damping[2,4]. We attempted to perform this analysis for our Bi$_2$Se$_3$/Py samples, but encountered the difficulty that the linewidth is not simple a linear function of the microwave frequency (Fig. S3) so that we cannot extract a meaningful measurement of the Gilbert damping parameter. We do not know the reason for this unusual behavior. One possible speculation (that we will pursue in future measurements) is that the interaction with the Bi$_2$Se$_3$ might cause spatially non-uniform magnetic configurations in the Py film that evolve as a function of applied magnetic field in the range corresponding to the data in Fig. S3.

## IV. Theoretical background and modeling

*Non-equilibrium spin accumulation of a topological surface state.* The topological surface state of Bi$_2$Se$_3$ can be described by the Dirac Hamiltonian $\mathcal{H}_{\vec{k}}^D = v_F \left( \hat{z} \times \vec{\sigma} \right) \cdot \vec{k} - \mu_D$ where $\vec{\sigma}$ are the Pauli matrices acting in spin space, $\hat{z}$ is the unit vector in the $z$ direction, and $v_F$ and $\mu_D$ are constants. This leads to the characteristic linear dispersion and spin helicity around the Fermi surface of the TI surface state. The velocity operator $\vec{v} = \partial_{\vec{k}} \mathcal{H}_{\vec{k}}^D$ is directly proportional to the spin operator $\vec{S} = (\hbar/2)\vec{\sigma}$ in the form $\vec{v} = 2v_F(\hat{z} \times \vec{S})/\hbar$. While the surface state has a vanishing equilibrium spin expectation value, any charge current flowing through the system leads to a nonzero spin accumulation independent of the microscopic details of the transport. A current density $j_x^D = en\langle v \rangle_{\text{neq}}$ in the $x$ direction (where $e$ is the electron's charge and $n$ is the electron density) produces a spin density in the $y$ direction

$$\langle S_y \rangle_{\text{neq}}^{D} = -\frac{\hbar}{2ev_F} j_x^D \ . \tag{S1}$$

*Spin accumulation due to the existence of additional non-topological surface states.* The Bi$_2$Se$_3$/ferromagnet interface is expected to not only host a topological Dirac state, but also a Rashba-split 2DEG with a dispersion $\mathcal{H}_{\vec{k}}^{R} = \frac{\vec{k}^2}{2m_R} + \alpha_R(\hat{z} \times \vec{\sigma}) \cdot \vec{k} - \mu_R$ (ref. 31). Figure S4a shows schematically the combined bandstructure for the case with band offset parameters $\mu_D > \mu_R > 0$. Both $v_F$ and $\alpha_R$ are positive, so that an alternating spin structure with a clockwise spin angular momentum on the TI surface state (corresponding to a counterclockwise spin moment) is obtained[31] (Fig. S4b).

The Rashba-split surface state leads to an additional spin accumulation $\langle S_y \rangle_{neq}^{R} = \frac{\hbar}{2e} \frac{m_R \alpha_R j_x^R}{2E_F}$ (ref. 32,33), where $E_F$ is the Fermi energy of the Rashba state. The spin accumulation of a Rashba-split surface state is partially compensated because the two sub-bands that are split by the spin-orbit coupling contribute to the spin accumulation with opposite signs, with the result that the spin accumulation from a Rashba-split surface state is smaller by a factor of $2\Delta k_F / k_F$ ($\Delta k_F$ is the splitting of the bands at $E_F$) compared to the Dirac state.

The total spin accumulation is

$$\langle S_y \rangle_{neq} = -\frac{\hbar}{2e} \left( \frac{j_x^D}{v_F} - \frac{m_R \alpha_R j_x^R}{2E_F} \right).$$

As long as $v_F$ and $\alpha_R$ have the same sign, the non-equilibrium spin accumulation due to the Rashba-split 2DEG and that due to the helical Dirac state are opposite. Based on the sign of the torque we measure experimentally, we can identify the topological surface state as the dominant source for the generation of torque in our experiments.

*Spin diffusion and torque generation.* In the following, we model the torque generation in Py through diffusion of the spin accumulation from the $Bi_2Se_3$ surface, Eq. (S1), into the Py. The diffusion (in the z direction) leads to a steady-state (itinerant) spin density determined by[34]

$$0 = -\vec{\nabla} \cdot \vec{\mathcal{J}}_i - \frac{1}{\tau_J}(\vec{S} \times \hat{m})_i - \frac{1}{\tau_\phi}[\hat{m} \times (\vec{S} \times \hat{m})]_i - \frac{S_i}{\tau_{sf}} , \quad \text{(S2)}$$

where the spin current (for the $i^{th}$ spin component) is given by $\vec{\mathcal{J}}_i = -\mathcal{D}\vec{\nabla} S_i$ with $\mathcal{D}$ the diffusion coefficient. The second term in Eq. (S2) describes the precession of the spins around the moment of the Py, with $\tau_J$ the spin precession time. The third term captures the relaxation of the spin component perpendicular to the Py magnetization direction $\hat{m}$, with $\tau_\phi$ the spin decoherence time, and the last term describes spin diffusion with time scale $\tau_{sf}$.

We solve Eq. (S2) subject to the requirement that there is no spin current through the outer boundary of the Py, *i.e.* $\mathcal{J}(d) = 0$, where $d$ is the thickness of the Py layer. For the Bi$_2$Se$_3$/Py interface, we assume that due to the exchange interaction the itinerant spins of the Py at the interface align with the spin density of the TI interface, *i.e.* $\vec{S}(0) = \chi \langle \vec{S} \rangle_{neq}$ with $\chi$ of order one (we will set $\chi = 1$ in the following). Note that this choice of the boundary condition for the diffusion equation is crucial. For the contrasting case of a pure spin-Hall-effect geometry, the torque would be due to a spin current injected into the ferromagnet, and thus the correct boundary condition would be a non-zero spin-current at the interface, *i.e.* at the interface $\langle \vec{S} \rangle_{neq} \approx 0$ and $\mathcal{J}(0) \neq 0$. For realistic parameters, $\tau_{sf} ? \tau_\phi, \tau_J$, this results in a spin-Hall torque that is almost completely in-plane.

For the boundary conditions corresponding to non-equilibrium spin accumulation at the Bi$_2$Se$_3$/Py interface, the solution for the spin distribution in the $z$ direction is

$$\hat{S}(z) = S_\parallel(z) + iS_\perp(z) = S_0 \frac{\cosh[k(z-d)]}{\cosh(kd)} \quad \text{(S3)}$$

with $k = (\lambda_{\parallel}^{-2} - i\lambda_J^{-2})^{1/2}$ and $\lambda_{\parallel}^{-2} = \lambda_{sf}^{-2} + \lambda_{\phi}^{-2}$, where $\lambda_J^2 = \mathcal{D}\tau_J$, $\lambda_{sf}^2 = \mathcal{D}\tau_{sf}$, and $\lambda_{\phi}^2 = \mathcal{D}\tau_{\phi}$. $S_{\parallel}(z)$ is the in-plane spin density and $S_0$ is the initial spin density (at $d = 0$), both of which are perpendicular to $\hat{m}$. The torque on the Py moments is given by the spatial change of the spin current compensated by the spin relaxation,

$$\hat{T} = \int_0^d dz \left[ -\partial_z \hat{\mathcal{J}}(z) - \frac{1}{\tau_{sf}} \hat{S}(z) \right],$$

where we again use the short forms $\hat{T} = T_{\parallel} + iT_{\perp}$ and $\hat{\mathcal{J}} = \mathcal{J}_{\parallel} + i\mathcal{J}_{\perp}$. Given the spin distribution in the $z$ direction of Eq. (S3) and using $d \to \infty$, we find

$$\hat{T} = S_0 \frac{\mathcal{D}}{k} (\frac{1}{\lambda_{\phi}^2} - \frac{i}{\lambda_J^2}).$$

For $\varphi = 45°$ (see Fig 2a), the spin polarization perpendicular to the magnetization of the Py is $1/\sqrt{2}$ of the total spin accumulation at the interface, Eq. (S1), and we find

$$\hat{T} = -\frac{\hbar}{2} \frac{\mathcal{D}}{k} (\frac{1}{\lambda_{\phi}^2} - \frac{i}{\lambda_J^2}) \frac{\sqrt{2}}{2} \frac{j_x}{ev_F}.$$

In analogy to the spin-Hall angle $\theta_{SH} = (2eJ_S)/(\hbar J)$, the spin torque angle for the Bi$_2$Se$_3$/ferromagnet interface is

$$\hat{\theta} = \frac{\hat{T}}{j_x} \frac{2e}{\hbar} = -\frac{\sqrt{2}}{2} \frac{\mathcal{D}}{v_F k} (\frac{1}{\lambda_{\phi}^2} - \frac{i}{\lambda_J^2}).$$

Using realistic numbers for the parameters: $\lambda_J = \lambda_{\phi} = 1$ nm, $\lambda_{sf} = 5$ nm (ref. 35), $v_F = 5 \times 10^5$ ms$^{-1}$, and a diffusion coefficient $\mathcal{D} \approx 10$ cm$^2$s$^{-1}$, we find in-plane and out-of-plane spin torque angles of order one. Figure S5 shows the ratio of out-of-plane to in-plane spin torque angle $\theta_{\perp}/\theta_{\parallel}$ as a function of $\lambda_J/\lambda_{\phi}$ to illustrate that for $\lambda_J : \lambda_{\phi}$ the two torque components are roughly the same size. Our experimental finding of a large spin torque angle for both in-plane and out-of-plane directions is thus consistently described as arising from the topological surface state of $Bi_2Se_3$.

**V. Estimate of a voltage signal arising from spin pumping and the inverse Edelstein effect.**

In analogy with calculations of spin pumping and the inverse spin Hall effect[36,37], spin pumping together with the inverse Edelstein effect should contribute to the ST-FMR measurement a signal with a symmetric lineshape and a magnitude

$$V_{sp} = \theta_\parallel \frac{e w \lambda_{BiSe} R}{2\pi} \tanh\left(\frac{t_{BiSe}}{2\lambda_{BiSe}}\right) \text{Re}(g_{\uparrow\downarrow}^{eff}) \left\langle \left(\vec{m} \times \dot{\vec{m}}\right)_x \right\rangle$$

Here $w$ is the sample width, $\lambda_{BiSe}$ is the length scale for the thickness of the $Bi_2Se_3$ involved in spin-charge conversion, $R$ is the sample resistance, $\varphi$ is the magnetic field angle, $t_{BiSe}$ is the thickness of the $Bi_2Se_3$ layer, and $\text{Re}(g_{\uparrow\downarrow}^{eff})$ is the real part of the effective spin mixing conductance. The sign of this contribution is opposite to the contribution to the symmetric part of the ST-FMR signal from the in-plane component of spin torque and AMR. The dependence on the angle of the applied magnetic field is expected to be the same as shown in Fig. 3b, so that the spin pumping and AMR contributions cannot be distinguished on this basis.

To estimate a rough upper bound on $|V_{sp}|$, we note that the value of $\lambda_{BiSe} \tanh(t_{BiSe} / 2\lambda_{BiSe})$ is always less than $t_{BiSe}/2$ so we will use this as an upper limit. For $\text{Re}(g_{\uparrow\downarrow}^{eff})$, we will use the value determined in Py/Pt samples[38], $\text{Re}(g_{\uparrow\downarrow}^{eff}) \approx 2 \times 10^{19}$ m$^{-2}$, although this might be a significant overestimate because the conductivity of Pt is much larger than $Bi_2Se_3$. For precession of a magnetic thin film with an out-of-plane demagnetization field $\mu_0 M_{eff}$, we calculate

$$\left\langle \left(\vec{m}\times\dot{\vec{m}}\right)_x \right\rangle = \omega\, \phi_p^2 \sin(\varphi) \sqrt{\frac{B_{ext}}{B_{ext}+\mu_0 M_{eff}}}$$

where $\phi_p$ is the maximum precession angle (in the sample plane). For the experimental conditions corresponding to the data in Fig. 3a of the main text, we can determine

$$\phi_p = \frac{1}{dR/d\varphi}\frac{2}{I_{RF}}\sqrt{\left(V_{mix}^S\right)^2+\left(V_{mix}^A\right)^2} \approx 0.0095,$$

where $V_{mix}^S$ and $V_{mix}^A$ are the coefficients of $F_S$ and $F_A$ determined by fitting the $V_{mix}$ signal to Eq. (2). With these assumptions and using $\theta_\parallel = 2.0\text{-}3.5$, we find an upper bound for the spin pumping voltage $V_{sp} \approx 5\text{-}9$ µV which is of the same order as the symmetric ST-FMR signal shown in Fig. 3a. Therefore, depending on the actual values of $\lambda_{BiSe}$ and $\text{Re}(g_{\uparrow\downarrow}^{eff})$ the spin pumping voltage might represent a significant correction to the ST-FMR analysis of the symmetric part of the ST-FMR resonance. However, to do so would in any case require a spin torque angle $\left|\theta_\parallel\right|$ much greater than 1, so that regardless of whether or not $V_{sp}$ is significant our conclusion that $Bi_2Se_3$ provides very large values of $\theta_\parallel$ and $\sigma_{S,\parallel}$ remains unchanged. Because spin pumping and the inverse Edelstein effect contribute only to the symmetric lineshape, any presence of these effects should not alter our interpretation of the antisymmetric part of the ST-FMR signals and the large value we determine for $\sigma_{S,\perp}$.

## VI. Calibration of the anisotropic magnetoresistance.

In Fig. S6, we show representative data demonstrating how we calibrate the angular dependence of the anisotropic magnetoresistance $R(\varphi)$ as input for the ST-FMR analysis.

**VII. Other spin-torque effects involving topological insulators.**

Other types of spin-torque effects involving TIs have also been proposed[39-43], but we do not believe that these are pertinent to our sample geometry and doping level.

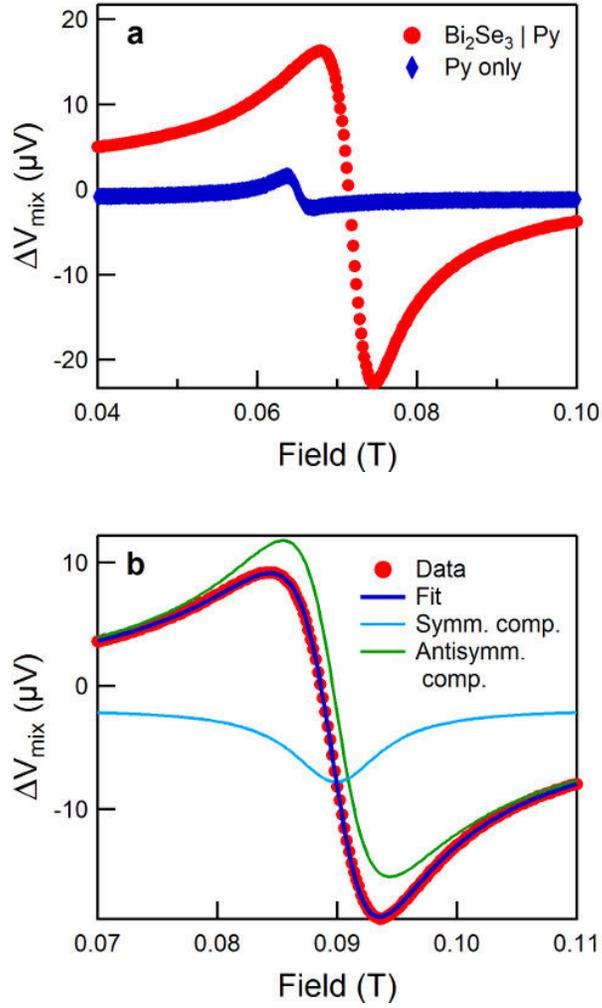

**Figure S1. a**, A comparison between the ST-FMR signals measured for two 50 μm × 15 μm samples at 8 GHz, one with 8 nm $Bi_2Se_3$/16 nm Py and the other a single layer of 16 nm Py. The absorbed RF power was 5 dBm and $\varphi = 45°$ in both cases. **b**, ST-FMR measurement for an 6 nm Pt/12 nm Py sample with dimensions 80 μm × 24 μm. The absorbed power was 5 dBm and $\varphi = 45°$.

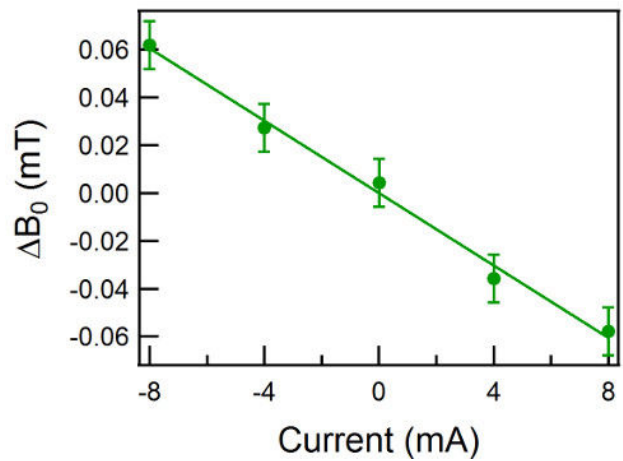

**Figure S2.** Change in resonant field as a function of direct current for an 8 nm $Bi_2Se_3$/16 nm Py device with dimensions 50 μm × 15 μm, averaged over frequencies between 6 and 10 GHz.

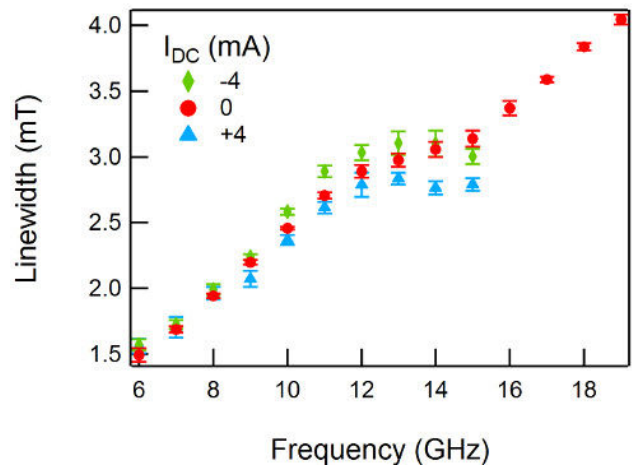

**Figure S3.** Measured linewidth verses frequency for an 8 nm $Bi_2Se_3$/16 nm Py device with dimensions 50 μm × 10 μm at different direct currents.

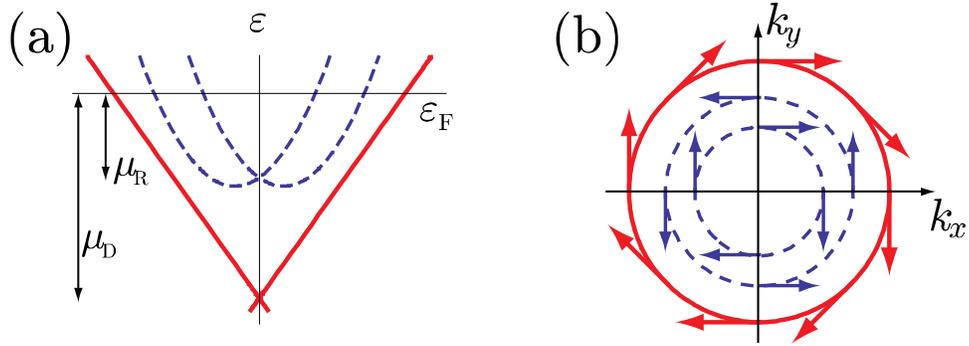

**Figure S4**. **a**, A schematic bandstructure with a Dirac surface state and a Rashba-split 2DEG as observed in references 26,31. **b**, The corresponding spin angular momentum structure at the Fermi energy.

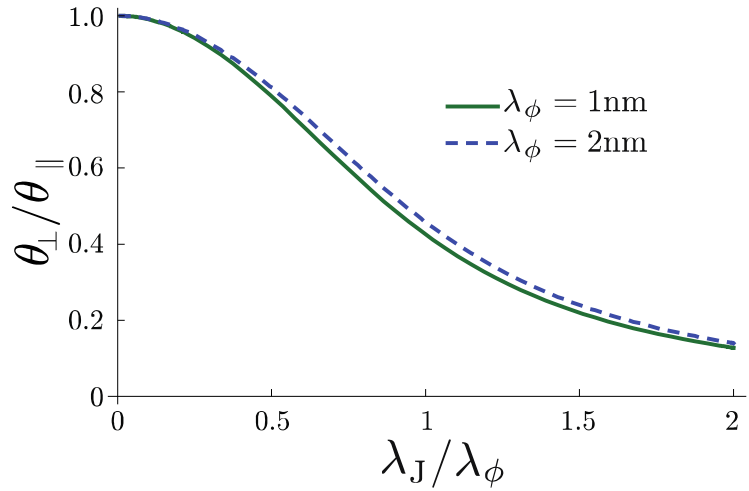

**Figure S5**. The ratio of the out-of-plane to the in-plane spin-torque angle $\theta_\perp / \theta_\parallel$ as a function of $\lambda_J / \lambda_\phi$, for $\lambda_{sf} = 5$ nm.

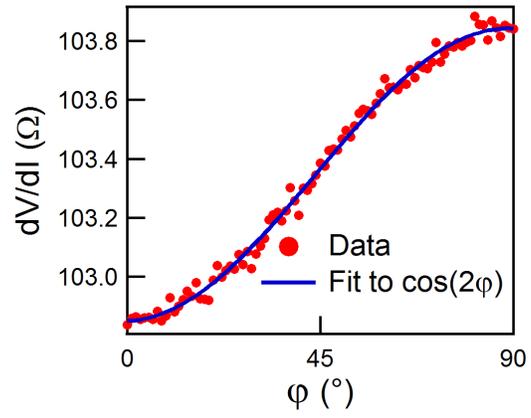

**Figure S6.** Anisotropic magnetoresistance calibration for an 8 nm $Bi_2Se_3$/16 nm Py sample at room temperature with dimensions 50 μm × 15 μm.

**Supplementary References**